# Dietary Supplements and Nutraceuticals Under Investigation for COVID-19 Prevention and Treatment

**This in progress manuscript is not intended for the general public.** This is a review paper that is authored by scientists for an audience of scientists to discuss research that is in progress. If you are interested in guidelines on testing, therapies, or other issues related to your health, you should not use this document. Instead, you should collect information from your local health department, the [CDC's guidance](CDC's guidance), or your own government.

# Authors

- **Ronan Lordan** [0000-0001-9668-3368](0000-0001-9668-3368) [RLordan](RLordan) [el_ronan](el_ronan) · Institute for Translational Medicine and Therapeutics, Perelman School of Medicine, University of Pennsylvania, Philadelphia, PA 19104-5158, USA

- **Halie M. Rando** [0000-0001-7688-1770](0000-0001-7688-1770) [rando2](rando2) [tamefoxtime](tamefoxtime) · Department of Systems Pharmacology and Translational Therapeutics, University of Pennsylvania, Philadelphia, Pennsylvania, United States of America; Department of Biochemistry and Molecular Genetics, University of Colorado School of Medicine, Aurora, Colorado, United States of America · Funded by the Gordon and Betty Moore Foundation (GBMF 4552)

- **COVID-19 Review Consortium**

- **Casey S. Greene** [0000-0001-8713-9213](0000-0001-8713-9213) [cgreene](cgreene) [GreeneScientist](GreeneScientist) · Department of Systems Pharmacology and Translational Therapeutics, University of Pennsylvania, Philadelphia, Pennsylvania, United States of America; Childhood Cancer Data Lab, Alex's Lemonade Stand Foundation, Philadelphia, Pennsylvania, United States of America; Department of Biochemistry and Molecular Genetics, University of Colorado School of Medicine, Aurora, Colorado, United States of America; Center for Health AI, University of Colorado School of Medicine, Aurora, Colorado, United States of America · Funded by the Gordon and Betty Moore Foundation (GBMF 4552); the National Human Genome Research Institute (R01 HG010067)

**COVID-19 Review Consortium:** Vikas Bansal, John P. Barton, Simina M. Boca, Christian Brueffer, James Brian Byrd, Stephen Capone, Shikta Das, Anna Ada Dattoli, John J. Dziak, Jeffrey M. Field, Soumita Ghosh, Anthony Gitter, Rishi Raj Goel, Casey S. Greene, Marouen Ben Guebila, Fengling Hu, Nafisa M. Jadavji, Sergey Knyazev, Likhitha Kolla, Alexandra J. Lee, Ronan Lordan, Tiago Lubiana, Temitayo Lukan, Adam L. MacLean, David Mai, Serghei Mangul,

David Manheim, Lucy D'Agostino McGowan, YoSon Park, Dimitri Perrin, Yanjun Qi, Diane N. Rafizadeh, Bharath Ramsundar, Halie M. Rando, Sandipan Ray, Michael P. Robson, Elizabeth Sell, Lamonica Shinholster, Ashwin N. Skelly, Yuchen Sun, Gregory L Szeto, Ryan Velazquez, Jinhui Wang, Nils Wellhausen

Authors with similar contributions are ordered alphabetically.

# Abstract

Coronavirus disease 2019 (COVID-19) has caused global disruption and a significant loss of life. Existing treatments that can be repurposed as prophylactic and therapeutic agents could reduce the pandemic's devastation. Emerging evidence of potential applications in other therapeutic contexts has led to the investigation of dietary supplements and nutraceuticals for COVID-19. Such products include vitamin C, vitamin D, omega 3 polyunsaturated fatty acids, probiotics, and zinc, all of which are currently under clinical investigation. In this review, we critically appraise the evidence surrounding dietary supplements and nutraceuticals for the prophylaxis and treatment of COVID-19. Overall, further study is required before evidence-based recommendations can be formulated, but nutritional status plays a significant role in patient outcomes, and these products could help alleviate deficiencies. For example, evidence indicates that vitamin D deficiency may be associated with greater incidence of infection and severity of COVID-19, suggesting that vitamin D supplementation may hold prophylactic or therapeutic value. A growing number of scientific organizations are now considering recommending vitamin D supplementation to those at high risk of COVID-19. Because research in vitamin D and other nutraceuticals and supplements is preliminary, here we evaluate the extent to which these nutraceutical and dietary supplements hold potential in the COVID-19 crisis.

# Importance

Sales of dietary supplements and nutraceuticals have increased during the pandemic due to their perceived "immune-boosting" effects. However, little is known about the efficacy of these dietary supplements and nutraceuticals against the novel coronavirus (SARS-CoV-2) or the disease it causes, COVID-19. This review provides a critical overview of the potential prophylactic and therapeutic value of various dietary supplements and nutraceuticals from the evidence available to date. These include vitamin C, vitamin D, and zinc, which are often perceived by the public as treating respiratory infections or supporting immune health. Consumers need to be aware of misinformation and false promises surrounding some supplements, which may be subject to limited regulation by authorities. However, considerably more research is required to determine whether dietary supplements and nutraceuticals exhibit prophylactic and therapeutic value against SARS-CoV-2 infection and COVID-19. This review provides perspective on which nutraceuticals and supplements are involved in biological processes that are relevant to recovery from or prevention of COVID-19.

# Introduction

The year 2020 saw scientists and the medical community scrambling to repurpose or discover novel host-directed therapies against the coronavirus diseases 2019 (COVID-19) pandemic caused by the spread of the novel *Severe acute respiratory syndrome-related coronavirus 2* (SARS-CoV-2). This rapid effort led to the identification of some promising pharmaceutical therapies for hospitalized patients, such as remdesivir and dexamethasone. Furthermore, most societies have adopted non-pharmacological preventative measures such as utilizing public health strategies that reduce the transmission of SARS-CoV-2. However, during this time, many individuals sought additional protections via the consumption of various dietary supplements and nutraceuticals that they believed to confer beneficial effects. While a patient's nutritional status does seem to play a role in COVID-19 susceptibility and outcomes [1,2,3,4,5], the beginning of the pandemic saw sales of vitamins and other supplements to soar despite a lack of any evidence supporting their use against COVID-19. In the United States, for example, dietary supplement and nutraceutical sales have shown modest annual growth in recent years (approximately 5%, or a $345 million increase in 2019), but during the six-week period preceding April 5, 2020, they increased by 44% ($435 million) relative to the same period in 2019 [6]. While growth subsequently leveled off, sales continued to boom, with a further 16% ($151 million) increase during the six weeks preceding May 17, 2020 relative to 2019 [6]. In France, New Zealand, India, and China, similar trends in sales were reported [7,8,9,10]. The increase in sales was driven by a consumer perception that dietary supplements and nutraceuticals would protect consumers from infection and/or mitigate the impact of infection due to the various "immune-boosting" claims of these products [11,12].

Due to the significant interest from the general public in dietary additives, whether and to what extent nutraceuticals or dietary supplements can provide any prophylactic or therapeutic benefit remains a topic of interest for the scientific community. Nutraceuticals and dietary supplements are related but distinct non-pharmaceutical products. Nutraceuticals are classified as supplements with health benefits beyond their basic nutritional value [13,14]. The key difference between a dietary supplement and a nutraceutical is that nutraceuticals should not only supplement the diet, but also aid in the prophylaxis and/or treatment of a disorder or disease [15]. However, dietary supplements and nutraceuticals, unlike pharmaceuticals, are not subject to the same regulatory protocols that protect consumers of medicines. Indeed, nutraceuticals do not entirely fall under the responsibility of the Food and Drug Administration (FDA), but they are monitored as dietary supplements according to the Dietary Supplement, Health and Education Act 1994 (DSHEA) [16] and the Food and Drug Administration Modernization Act 1997 (FDAMA) [17]. Due to increases in sales of dietary supplements and nutraceuticals, in 1996 the FDA established the Office of Dietary Supplement Programs (ODSP) to increase surveillance. Novel products or nutraceuticals must now submit a new dietary ingredient notification to the ODSP for review. There are significant concerns that these legislations do not adequately protect the consumer as they ascribe responsibility to the manufacturers to ensure the safety of the product before manufacturing or marketing [18]. Manufacturers are not required to register or even seek approval from the FDA to produce or sell food supplements or nutraceuticals. Health or nutrient content claims for labeling purposes are approved based on an authoritative statement from the Academy of Sciences or relevant federal authorities once the FDA has been notified and on the basis that the information is known to be true and not deceptive [18].

Therefore, there is often a gap between perceptions by the American public about a nutraceutical or dietary supplement and the actual clinical evidence surrounding its effects.

Despite differences in regulations, similar challenges exist outside of the United States. In Europe, where the safety of supplements are monitored by the European Union (EU) under Directive 2002/46/EC [19]. However, nutraceuticals are not directly mentioned. Consequently, nutraceuticals can be generally described as either a medicinal product under Directive 2004/27/EC [20] or as a 'foodstuff' under Directive 2002/46/EC of the European council. In order to synchronize the various existing legislations, Regulation EC 1924/2006 on nutrition and health claims was put into effect to assure customers of safety and efficacy of products and to deliver understandable information to consumers. However, specific legislation for nutraceuticals is still elusive. Health claims are permitted on a product label only following compliance and authorization according to the European Food Safety Authority (EFSA) guidelines on nutrition and health claims [21]. EFSA does not currently distinguish between food supplements and nutraceuticals for health claim applications of new products, as claim authorization is dependent on the availability of clinical data in order to substantiate efficacy [22]. These guidelines seem to provide more protection to consumers than the FDA regulations but potentially at the cost of innovation in the sector [23]. The situation becomes even more complicated when comparing regulations at a global level, as countries such as China and India have existing regulatory frameworks for traditional medicines and phytomedicines not commonly consumed in Western society [24]. Currently, there is debate among scientists and regulatory authorities surrounding the development of a widespread regulatory framework to deal with the challenges of safety and health claim substantiation for nutraceuticals [18,22], as these products do not necessarily follow the same rigorous clinical trial frameworks used to approve the use of pharmaceuticals. Such regulatory disparities have been highlighted by the pandemic, as many individuals and companies have attempted to profit from the vulnerabilities of others by overstating claims in relation to the treatment of COVID-19 using supplements and nutraceuticals. The FDA has written several letters to prevent companies marketing or selling products based on false hyperbolic promises about preventing SARS-CoV-2 infection or treating COVID-19 [25,26,27]. These letters came in response to efforts to market nutraceutical prophylactics against COVID-19, some of which charged the consumer as much as $23,000 [28]. There have even been some incidents highlighted in the media because of their potentially life threatening consequences; for example, the use of oleandrin was touted as a potential "cure" by individuals close to the former President of the United States despite its high toxicity [29]. Thus, heterogeneous and at times relaxed regulatory standards have permitted high-profile cases of the sale of nutraceuticals and dietary supplements that are purported to provide protection against COVID-19, despite a lack of research into these compounds.

Notwithstanding the issues of poor safety, efficacy, and regulatory oversight, some dietary supplements and nutraceuticals have exhibited therapeutic and prophylactic potential. Some have been linked with reduced immunopathology, antiviral and anti-inflammatory activities, or even the prevention of acute respiratory distress syndrome (ARDS) [11,30,31]. A host of potential candidates have been highlighted in the literature that target various aspects of the COVID-19 viral pathology, while others are thought to prime the host immune system. These candidates include vitamins and minerals along with extracts and omega-3 polyunsaturated fatty acids (n-3 PUFA) [32]. *In vitro* and *in vivo* studies suggest that nutraceuticals containing phycocyanobilin, N-acetylcysteine, glucosamine, selenium or phase 2 inductive nutraceuticals (e.g. ferulic acid, lipoic acid, or sulforaphane) can prevent or modulate RNA virus infections via amplification of the signaling activity of mitochondrial antiviral-signaling protein (MAVS) and activation of Toll-like receptor 7 [33]. While promising, further animal and human studies are required to assess the therapeutic potential of these various nutrients and nutraceuticals against COVID-19. For the purpose of this review, we have highlighted some of the main dietary supplements and nutraceuticals that are currently under investigation for their potential prophylactic and therapeutic applications. These include n-3 PUFA, zinc, vitamins C and D, and probiotics.

## n-3 PUFA

One category of supplements that has been explored for beneficial effects against various viral infections are the n-3 PUFAs [32], commonly referred to as omega-3 fatty acids, which include eicosapentaenoic acid (EPA) and docosahexaenoic acid (DHA). EPA and DHA intake can come from a diet high in fish or through dietary supplementation with fish oils or purified oils [34]. Other, more sustainable sources of EPA and DHA include algae [35,36], which can also be exploited for their rich abundance of other bioactive compounds such as angiotensin converting enzyme inhibitor peptides and antiviral agents including phycobiliproteins, sulfated polysaccharides, and calcium-spirulan [37]. n-3 PUFAs have been investigated for many years for their therapeutic potential [38]. Supplementation with fish oils is generally well tolerated [38], and intake of n-3 PUFAs through dietary sources or supplementation is specifically encouraged for vulnerable groups such as pregnant and lactating women [39,40]. As a result, these well-established compounds have drawn significant interest for their potential immune effects and therapeutic potential.

Particular interest has arisen in n-3 PUFAs as potential therapeutics against diseases associated with inflammation. n-3 PUFAs have been found to mediate inflammation by influencing processes such as leukocyte chemotaxis, adhesion molecule expression, and the production of eicosanoids [41,42]. This and other evidence indicates that n-3 PUFAs may have the capacity to modulate the adaptive immune response [14,34,41]; for example, they have been found to influence antigen presentation and the production of CD4(+) Th1 cells, among other relevant effects [43]. Certainly, preliminary evidence from banked blood samples from 100 COVID-19 patients suggests that patients with a higher omega-3 index, a measure of n-3 and n-6 fatty acids in red blood cells, had a lower risk of death due to COVID-19 [44]. Interest has also arisen as to whether nutritional status related to n-3 PUFAs can also affect inflammation associated with severe disease, such as ARDS or sepsis [45,46]. ARDS and sepsis hold particular concern in the treatment of severe COVID-19; an analysis of 82 deceased COVID-19 patients in Wuhan during January to February 2020 reported that respiratory failure (associated with ARDS) was the cause of death in 69.5% of cases, and sepsis or multi-organ failure accounted for 28.0% of deaths [47]. Research in ARDS prior to current pandemic suggests that n-3 PUFAs may hold some therapeutic potential. One study randomized 16 consecutive ARDS patients to receive either a fish oil-enriched lipid emulsion or a control lipid emulsion (comprised of 100% long-chain triglycerides) under a double-blinded design [48]. They reported a statistically significant reduction in leukotriene B4 levels in the group receiving the fish oil-enriched emulsion, suggesting that the fish oil supplementation may have reduced inflammation. However, they also reported that most of their tests were not statistically significant, and therefore it seems that additional research using larger sample sizes is required. A recent meta-analysis of 10 randomized controlled trials (RCTs) examining the effects of n-3 PUFAs on ARDS patients did not find evidence of any effect on mortality, although the effect on secondary outcomes could not be determined due to a low quality of evidence [49]. However, another meta-analysis that examined 24 RCTs studying the effects of n-3 fatty acids on sepsis, including ARDS-induced sepsis, did find support for an effect on mortality when n-3 fatty acids were administered via enteral nutrition, although a paucity of high-quality evidence again limited conclusions [50]. Therefore, despite theoretical support for an immunomodulatory effect of n-3 PUFAs in COVID-19, evidence from existing RCTs is insufficient to determine whether supplementation offers an advantage in a clinical setting that would be relevant to COVID-19.

Another potential mechanism that has led to interest in n-3 PUFAs as protective against viral infections including COVID-19 is its potential as a precursor molecule for the biosynthesis of endogenous specialized proresolving mediators (SPM), such as protectins and resolvins, that actively resolve inflammation and infection [51]. SPM have exhibited beneficial effects against a variety of lung infections, including some caused by RNA viruses [52,53]. Indeed, protectin D1 has been shown to increase survival from H1N1 viral infection in mice by affecting the viral replication machinery [54]. Several mechanisms for SPM have been proposed, including preventing the release of pro-inflammatory cytokines and chemokines or increasing phagocytosis of cellular debris by macrophages [55]. In influenza, SPM promote antiviral B lymphocytic activities [56], and protectin D1 has been shown to increase survival from H1N1 viral infection in mice by affecting the viral replication machinery [54]. It has thus been hypothesized that SPM could aid in the resolution of the cytokine storm and pulmonary inflammation associated with COVID-19 [57,58]. Another theory is that some comorbidities, such as obesity, could lead to deficiencies of SPM, which could in turn be related to the occurrence of adverse outcomes for COVID-19 [59]. However, not all studies are in agreement that n-3 PUFAs or their resulting SPM are effective against infections [60]. At a minimum, the effectiveness of n-3 PUFAs against infections would be dependent on the dosage, timing, and the specific pathogens responsible [61]. On another level, there is still the question of whether fish oils can raise the levels of SPM levels upon ingestion and in response to acute inflammation in humans [62]. Currently, Karolinska University Hospital is running a trial that will measure the levels of SPM as a secondary outcome following intravenous supplementation of n-3 PUFAs in hospitalized COVID-19 patients to determine whether n-3 PUFAs provides therapeutic value [63,64]. Therefore, while this mechanism provides theoretical support for a role for n-3 PUFAs against COVID-19, experimental support is still needed.

A third possible mechanism by which n-3 PUFAs could benefit COVID-19 patients arises from the fact that some COVID-19 patients, particularly those with comorbidities, are at a significant risk of thrombotic complications including arterial and venous thrombosis [65,66]. Therefore, the use of prophylactic and therapeutic anticoagulants and antithrombotic agents is under consideration [67,68]. Considering that there is significant evidence that n-3 fatty acids and other fish oil-derived lipids possess antithrombotic properties and anti-inflammatory properties [34,69,70], they may have therapeutic value against the prothrombotic complications of COVID-19. In particular, concerns have been raised within the medical community about using investigational therapeutics on COVID-19 patients who are already on antiplatelet therapies due to pre-existing comorbidities because the introduction of such therapeutics could lead to issues with dosing and drug choice and/or negative drug-drug interactions [67]. In such cases, dietary sources of n-3 fatty acids or other nutraceuticals with antiplatelet activities could hold particular value for reducing the risk of thrombotic complications in patients already receiving pharmaceutical antiplatelet therapies. A new clinical trial [71] is currently recruiting COVID-19 positive patients to investigate the anti-inflammatory activity of a recently developed, highly purified nutraceutical derivative of EPA known as icosapent ethyl (Vascepa™) [72]. Other randomized controlled trials that are in the preparatory stages intend to investigate the administration of EPA and other bioactive compounds to COVID-19 positive patients in order to observe whether anti-inflammatory effects or disease state improvements occur [73,74]. Finally, while there have been studies investigating the therapeutic value of n-3 fatty acids against ARDS in humans, there is still limited evidence of their effectiveness [75]. It should be noted that the overall lack of human studies in this area means there is limited evidence as to whether these supplements could affect COVID-19 infection. Consequently, the clinical trials that are underway and those that have been proposed will provide valuable insight into whether the anti-inflammatory potential of n-3 PUFAs and their derivatives can be beneficial to the treatment of COVID-19. All the same, while the evidence is not present to draw conclusions about whether n-3 PUFAs will be useful in treating COVID-19, there is likely little harm associated with a diet rich in fish oils, and interest in n-3 PUFA supplementation by the general public is unlikely to have negative effects.

## Zinc

Zinc is nutrient supplement that may exhibit some benefits against RNA viral infections. Zinc is a trace metal obtained from dietary sources or supplementation and is important for the maintenance of immune cells involved in adaptive and innate immunity [76]. Supplements can be administered orally as a tablet or as a lozenge and are available in many forms, such as zinc picolinate, zinc acetate, and zinc citrate. Zinc is also available from dietary sources including meat, seafood, nuts, seeds, legumes, and dairy. The role of zinc in immune function has been extensively reviewed [76]. Zinc is an important signaling molecule, and zinc levels can alter host defense systems. In inflammatory situations such as an infection, zinc can regulate leukocyte immune responses and activate the nuclear factor kappa-light-chain-enhancer of activated B cells, thus altering cytokine production [77,78]. In particular, zinc supplementation can increase natural killer cell levels, which are important cells for host defense against viral infections [76,79]. As a result of these immune-related functions, zinc is also under consideration for possible benefits against COVID-19.

Adequate zinc intake has been associated with reduced incidence of infection [80] and antiviral immunity [81]. A randomized, double-blind, placebo-controlled trial that administered zinc supplementation to elderly subjects over the course of a year found zinc deficiency to be associated with increased susceptibility to infection and that zinc deficiency could be prevented through supplementation [80]. Clinical trial data supports the utility of zinc to diminish the duration and severity of symptoms associated with common colds when it is provided within 24 hours of the onset of symptoms [82,83]. An observational study showed that COVID-19 patients had significantly lower zinc levels in comparison to healthy controls and that zinc-deficient COVID-19 patients (those with levels less than 80 μg/dl) tended to have more complications

(70.4% vs 30.0%, *p* = 0.009) and potentially prolonged hospital stays (7.9 vs 5.7 days, *p* = 0.048) relative to patients who were not zinc deficient [84]. In coronaviruses specifically, *in vitro* evidence has demonstrated that the combination of zinc ($Zn^{2+}$) and zinc ionophores (pyrithione) can interrupt the replication mechanisms of SARS-CoV-GFP (a fluorescently tagged SARS-CoV-1) and a variety of other RNA viruses [85,86]. Currently, there are over twenty clinical trials registered with the intention to use zinc in a preventative or therapeutic manner for COVID-19. However, many of these trials proposed the use of zinc in conjunction with hydroxychloroquine and azithromycin [87,88,89,90], and it is not known how the lack of evidence supporting the use of hydroxychloroquine will affect investigation of zinc. One retrospective observational study of New York University Langone hospitals in New York compared outcomes among hospitalized COVID-19 patients administered hydroxychloroquine and azithromycin with zinc sulfate (n = 411) versus hydroxychloroquine and azithromycin alone (n = 521). Notably, zinc is the only treatment that was used in this trial that is still under consideration as a therapeutic agent due to to the lack of efficacy and potential adverse events associated with hydroxychloroquine and azithromycin against COVID-19 [91,92,93]. While the addition of zinc sulfate did not affect the duration of hospitalization, the length of ICU stays or patient ventilation duration, univariate analyses indicated that zinc did increase the frequency of patients discharged and decreased the requirement for ventilation, referrals to the ICU, mortality [94]. However, a smaller retrospective study at Hoboken University Medical Center New Jersey failed to find an association between zinc supplementation and survival of hospitalized patients [95]. Therefore, whether zinc contributes to COVID-19 recovery remains unclear. Other trials are now investigating zinc in conjunction with other supplements such as vitamin C or n-3 PUFA [74,96]. Though there is, overall, encouraging data for zinc supplementation against the common cold and viral infections, there is currently limited evidence to suggest zinc supplementation has any beneficial effects against the current novel COVID-19; thus, the clinical trials that are currently underway will provide vital information on the efficacious use of zinc in COVID-19 prevention and/or treatment. However, given the limited risk and the potential association between zinc deficiency and illness, maintaining a healthy diet to ensure an adequate zinc status may be advisable for individuals seeking to reduce their likelihood of infection.

## Vitamin C

Vitamins B, C, D, and E have also been suggested as potential nutrient supplement interventions for COVID-19 [32,97]. In particular vitamin C has been proposed as a potential therapeutic agent against COVID-19 due to its long history of use against the common cold and other respiratory infections [98,99]. Vitamin C can be obtained via dietary sources such as fruits and vegetables or via supplementation. Vitamin C plays a significant role in promoting immune function due to its effects on various immune cells. It affects inflammation by modulating cytokine production, decreasing histamine levels, enhancing the differentiation and proliferation of T- and B-lymphocytes, increasing antibody levels, and protecting against the negative effects of reactive oxygen species, among other effects related to COVID-19 pathology [100,101,102]. Vitamin C is utilized by the body during viral infections, as evinced by lower concentrations in leukocytes and lower concentrations of urinary vitamin C. Post-infection, these levels return to baseline ranges [103,104,105,106,107]. It has been shown that as little as 0.1 g/d of vitamin C can maintain normal plasma levels of vitamin C in healthy individuals, but higher doses of at least 1-3 g/d are required for critically ill patients in ICUs [108]. Indeed, vitamin C deficiency appears to be common among COVID-19 patients [109,110]. COVID-19 is also associated with the formation of microthrombi and coagulopathy [111] that contribute to its characteristic lung pathology [112], but these symptoms can be ameliorated by early infusions of vitamin C to inhibit endothelial surface P-selectin expression and platelet-endothelial adhesion [113]. Intravenous vitamin C also reduced D-dimer levels, which are notably elevated in COVID-19 patients [114,115], in a case study of 17 COVID-19 patients [116]. There is therefore preliminary evidence suggesting that vitamin C status and vitamin C administration may be relevant to COVID-19 outcomes.

Larger-scale studies of vitamin C, however, have provided mixed results. A recent meta-analysis found consistent support for regular vitamin C supplementation reducing the duration of the common cold, but that supplementation with vitamin C (> 200 mg) failed to reduce the incidence of colds [117]. Individual studies have found Vitamin C to reduce the susceptibility of patients to lower respiratory tract infections, such as pneumonia [118]. Another meta-analysis demonstrated that in twelve trials, vitamin C supplementation reduced the length of stay of patients in intensive care units (ICUs) by 7.8% (95% CI: 4.2% to 11.2%; $p$ = 0.00003). Furthermore, high doses (1-3 g/day) significantly reduced the length of an ICU stay by 8.6% in six trials ($p$ = 0.003). Vitamin C also shortened the duration of mechanical ventilation by 18.2% in three trials in which patients required intervention for over 24 hours (95% CI 7.7% to 27%; p = 0.001) [108]. Despite these findings, an RCT of 167 patients known as CITRUS ALI failed to show a benefit of a 96-hour infusion of vitamin C to treat ARDS [119]. Clinical trials specifically investigating vitamin C in the context of COVID-19 have now begun, as highlighted by Carr et al. [99]. These trials intend to investigate the use of intravenous vitamin C in hospitalized COVID-19 patients. The first trial to report initial results took place in Wuhan, China [120]. These initial results indicated that the administration of 12 g/12 hr of intravenous vitamin C for 7 days in 56 critically ill COVID-19 patients resulted in a promising reduction of 28-day mortality ($p$ = 0.06) in univariate survival analysis [121]. Indeed, the same study reported a significant decrease in IL-6 levels by day 7 of vitamin C infusion ($p$ = 0.04) [122]. Additional studies that are being conducted in Canada, China, Iran, and the USA will provide additional insight into whether vitamin C supplementation affects COVID-19 outcomes on a larger scale.

Even though evidence supporting the use of vitamin C is beginning to emerge, we will not know how effective vitamin C is as a therapeutic for quite some time. Currently (as of January 2021) over fifteen trials are registered with clinicaltrials.gov that are either recruiting, active or are currently in preparation. When completed, these trials will provide crucial evidence on the efficacy of vitamin C as a therapeutic for COVID-19 infection. However, the majority of supplementation studies investigate the intravenous infusion of vitamin C in severe patients. Therefore, there is a lack of studies investigating the potential prophylactic administration of vitamin C via oral supplementation for healthy individuals or potentially asymptomatic SARS-CoV-2 positive patients. Once again, vitamin C intake is part of a healthy diet and the vitamin likely presents minimal risk, but its potential prophylactic or therapeutic effects against COVID-19 are yet to be determined. To maintain vitamin C status, it would be prudent for individuals to ensure that they consume the recommended dietary allowance of vitamin C to maintain a healthy immune system [1]. The recommended dietary allowance according to the FDA is 75-90 mg/d, whereas EFSA recommends 110 mg/d [123].

# Vitamin D

Of all of the supplements currently under investigation, vitamin D has become a leading prophylactic and therapeutic candidate against SARS-CoV-2. Vitamin D can modulate both the adaptive and innate immune system and is associated with various aspects of immune health and antiviral defense [124,125,126,127,128]. Vitamin D can be sourced through diet or supplementation, but it is mainly biosynthesized by the body on exposure to ultraviolet light (UVB) from sunlight. Vitamin D deficiency is associated with an increased susceptibility to infection [129]. In particular, vitamin D deficient patients are at risk of developing acute respiratory infections [130] and ARDS [130]. 1,25-dihydroxyvitamin D3 is the active form of vitamin D that is involved in adaptive and innate responses; however, due to its low concentration and a short half life of a few hours, vitamin D levels are typically measured by the longer lasting and more abundant precursor 25-hydroxyvitamin D. The vitamin D receptor is expressed in various immune cells, and vitamin D is an immunomodulator of antigen presenting cells, dendritic cells, macrophages, monocytes, and T- and B-lymphocytes [129,131]. Due to its potential immunomodulating properties, vitamin D supplementation may be advantageous to maintain a healthy immune system.

Early in the pandemic it was postulated that an individual's vitamin D status could significantly affect their risk of developing COVID-19 [132]. This hypothesis was derived from the fact that the current pandemic emerged in Wuhan China during winter, when 25-hydroxyvitamin D concentrations are at their lowest due to a lack of sunlight, whereas in the Southern Hemisphere, where it was nearing the end of the summer and higher 25-hydroxyvitamin D concentrations would be higher, the number of cases was low. This led researchers to question whether there was a seasonal component to the SARS-CoV-2 pandemic and whether vitamin D levels might play a role [132,133,134,135]. Though it is assumed that COVID-19 is seasonal, multiple other factors that can affect vitamin D levels should also be considered. These factors include an individual's nutritional status, their age, their occupation, skin pigmentation, potential comorbidities, and the variation of exposure to sunlight due to latitude amongst others. Indeed, it has been estimated that each degree of latitude north of 28 degrees corresponded to a 4.4% increase of COVID-19 mortality, indirectly linking a persons vitamin D levels via exposure to UVB light to COVID-19 mortality [133].

As the pandemic has evolved, additional research of varying quality has investigated some of the potential links identified early in the pandemic [132] between vitamin D and COVID-19. Indeed, studies are beginning to investigate whether there is any prophylactic and/or therapeutic relationship between vitamin D and COVID-19. A study in Switzerland demonstrated that 27 SARS-CoV-2 positive patients exhibited 25-hydroxyvitamin D plasma concentrations that were significantly lower (11.1 ng/ml) than those of SARS-CoV-2 negative patients (24.6 ng/ml; $p$ = 0.004), an association that held when stratifying patients greater than 70 years old [136]. These findings seem to be supported by a Belgian observational study of 186 SARS-CoV-2 positive patients exhibiting symptoms of pneumonia, where 25-hydroxyvitamin D plasma concentrations were measured and CT scans of the lungs were obtained upon hospitalization [137]. A significant difference in 25-hydroxyvitamin D levels was observed between the SARS-CoV-2 patients and 2,717 season-matched diseased controls. Both female and male patients possessed lower median 25-hydroxyvitamin D concentrations than the control group as a whole

(18.6 ng/ml versus 21.5 ng/ml; *p* = 0.0016) and a higher rate of vitamin D deficiency (58.6% versus 42.5%). However, when comparisons were stratified by sex, evidence of sexual dimorphism became apparent, as female patients had equivalent levels of 25-hydroxyvitamin D to females in the control group, whereas male patients were deficient in 25-hydroxyvitamin D relative to male controls (67% versus 49%; *p* = 0.0006). Notably, vitamin D deficiency was progressively lower in males with advancing radiological disease stages (*p* = 0.001). These studies are supported by several others that indicate that vitamin D status may be an independent risk factor for the severity of COVID-19 [138,139,140,141] and in COVID-19 patients relative to population-based controls [142]. Indeed, serum concentrations of 25-hydroxyvitamin D above 30 ng/ml, which indicate vitamin D sufficiency, seems to be associated with a reduction in serum C-reactive protein, an inflammatory marker, along with increased lymphocyte levels, which suggests that vitamin D levels may modulate the immune response by reducing risk for cytokine storm in response to SARS-CoV-2 infection [142]. A study in India determined that COVID-19 fatality was higher in patients with severe COVID-19 and low serum 25-hydroxyvitamin D (mean level 6.2 ng/ml; 97% vitamin D deficient) levels versus asymptomatic non-severe patients with higher levels of vitamin D (mean level 27.9 ng/ml; 33% vitamin D deficient) [143]. In the same study, vitamin D deficiency was associated with higher levels of inflammatory markers including IL-6, ferritin, and tumor necrosis factor α. Collectively, these studies add to a multitude of observational studies reporting potential associations between low levels of 25-hydroxyvitamin D and COVID-19 incidence and severity [136,141,142,144,145,146,147,148,149,150].

Despite the large number of studies establishing a link between vitamin D status and COVID-19 severity, an examination of data from the UK Biobank did not support this thesis [151,152]. These analyses examined 25-hydroxyvitamin D concentrations alongside SARS-CoV-2 positivity and COVID-19 mortality in over 340,000 UK Biobank participants. However, these studies have caused considerable debate that will likely be settled following further studies [153,154]. Overall, while the evidence suggests that there is likely an association between low serum 25-hydroxyvitamin D and COVID-19 incidence, these studies must be interpreted with caution, as there is the potential for reverse causality, bias, and other confounding factors including that vitamin D deficiency is also associated with numerous pre-existing conditions and risk factors that can increase the risk for severe COVID-19 [1,133,155,156].

While these studies inform us of the potential importance of vitamin D sufficiency and the risk of SARS-CoV-2 infection and severe COVID-19, they fail to conclusively determine whether vitamin D supplementation can therapeutically affect the clinical course of COVID-19. In one study, 40 vitamin D deficient asymptomatic or mildly symptomatic participants patients were either randomized to receive 60,000 IU of cholecalciferol daily for at least 7 days (n = 16) or a placebo (n = 24) with a target serum 25-hydroxyvitamin D level >50 ng/ml. At day 7, 10 patients achieved >50 ng/ml, followed by another 2 by day 14. By the end of the study, the treatment group had a greater proportion of vitamin D-deficient participants that tested negative for SARS-CoV-2 RNA, and they had a significantly lower fibrinogen levels, potentially indicating a beneficial effect [157]. A pilot study in Spain determined that early administration of high dose calcifediol (~21,000 IU days 1-2 and ~11,000 IU days 3-7 of hospital admission) with hydroxychloroquine and azithromycin to 50 hospitalized COVID-19 patients significantly reduced ICU admissions and may have reduced disease severity versus hydroxychloroquine and azithromycin alone [158]. Although this study received significant criticism from the National Institute for Health and Care Excellence (NICE) in the UK [159], an independent follow-up statistical analysis supported the findings of the study with respect to the results of cholecalciferol treatment [160]. Another trial of 986 patients hospitalized for COVID-19 in three UK hospitals administered cholecalciferol supplementation (≥ 280,000 IU in a time period of 7 weeks) to 151 patients and found an association with a reduced risk of COVID-19 mortality, regardless of baseline 25-hydroxyvitamin D levels [161]. However, a double-blind, randomized, placebo-controlled trial of 240 hospitalized COVID-19 patients in São Paulo, Brazil administered a single 200,000 IU oral dose of vitamin D. While levels of 25-hydroxyvitamin D did increase from 21% to ~27% ($p$ = 0.001) and the supplementation was well tolerated, there was no reduction in the length of hospital stay or mortality and no change to any other relevant secondary outcomes [162]. These early findings are thus still inconclusive with regards to the therapeutic value of vitamin D supplementation. However, other trials are underway, including one trial that is investigating the utility of vitamin D as an immune-modulating agent by monitoring whether administration of vitamin D precipitates an improvement of health status in non-severe symptomatic COVID-19 patients and whether vitamin D prevents patient deterioration [163]. Other trials are examining various factors including mortality, symptom recovery, severity of disease, rates of ventilation, inflammatory markers such as C-reactive protein and IL-6, blood cell counts, and the prophylactic capacity of vitamin D administration [163,164,165,166]. Concomitant administration of vitamin D with pharmaceuticals such as aspirin [167] and bioactive molecules such as resveratrol [168] are also under investigation.

The effectiveness of vitamin D supplementation against COVID-19 remains open for debate. All the same, there is no doubt that vitamin D deficiency is a widespread issue and should be addressed not only because of its potential link to SARS-CoV-2 incidence [169], but also due to its importance for overall health. There is a possibility that safe exposure to sunlight could improve endogenous synthesis of vitamin D, potentially strengthening the immune system. However, sun exposure is not sufficient on its own, particularly in the winter months. Indeed, while the possible link between vitamin D status and COVID-19 is further investigated, preemptive supplementation of vitamin D and encouraging people to maintain a healthy diet for optimum vitamin D status is likely to raise serum levels of 25-hydroxyvitamin D while being unlikely to carry major health risks. These principles seem to be the basis of a number of guidelines issued by some countries and scientific organizations that have advised supplementation of vitamin D during the pandemic. The Académie Nationale de Médecine in France recommends rapid testing of 25-hydroxyvitamin D for people over 60 years old to identify those most at risk of vitamin D deficiency and advises them to obtain a bolus dose of 50,000 to 100,000 IU vitamin D to limit respiratory complications. It has also recommended that those under 60 years old should take 800 to 1,000 IU daily if they receive a SARS-CoV-2 positive test [170]. In Slovenia, doctors have been advised to provide nursing home patients with vitamin D [171]. Both Public Health England and Public Health Scotland have advised members of the Black, Asian, and minority ethnic communities to supplement for vitamin D in light of evidence that they may be at higher risk for vitamin D deficiency along with other COVID-19 risk factors, a trend that has also been observed in the United States [172,173]. However, other UK scientific bodies including the NICE recommend that individuals supplement for vitamin D as per usual UK government advice but warn that people should not supplement for vitamin D solely to prevent COVID-19. All the same, the NICE has provided guidelines for research to investigate the supplementation of vitamin D in the context of COVID-19 [174]. Despite vitamin D deficiency being a widespread issue in the United States [175], the National Institutes of Health have stated that there is "insufficient data to recommend either for or against the use of vitamin D for the prevention or treatment of COVID-19" [176]. These are just some examples of how public health guidance has responded to the emerging evidence regarding vitamin D and COVID-19. Outside of official recommendations, there is also evidence that individuals may be paying increased attention to their vitamin D levels, as a survey of Polish consumers showed that 56% of respondents used vitamin D during the pandemic [177]. However, some companies have used the emerging evidence surrounding vitamin D to sell products that claim to prevent and treat COVID-19, which in one incident required a federal court to intervene and issue an injunction barring the sale of vitamin-D-related products due to the lack of clinical data supporting these claims [178]. It it clear that further studies and clinical trials are required to conclusively determine the prophylactic and therapeutic potential of vitamin D supplementation against COVID-19. Until such time that sufficient evidence emerges, individuals should follow their national guidelines surrounding vitamin D intake to achieve vitamin D sufficiency.

## Probiotics

Probiotics are "live microorganisms that, when administered in adequate amounts, confer a health benefit on the host" [179]. Some studies suggest that probiotics are beneficial against common viral infections, and there is modest evidence to suggest that they can modulate the immune response [180,181]. As a result, it has been hypothesized that probiotics may have therapeutic value worthy of investigation against SARS-CoV-2 [182]. Probiotics and next-generation probiotics, which are more akin to pharmacological-grade supplements, have been associated with multiple potential beneficial effects for allergies, digestive tract disorders, and even metabolic diseases through their anti-inflammatory and immunomodulatory effects [183,184]. However, the mechanisms by which probiotics affect these various conditions would likely differ among strains, with the ultimate effect of the probiotic depending on the heterogeneous set of bacteria present [184]. Some of the beneficial effects of probiotics include reducing inflammation by promoting the expression of anti-inflammatory mediators, inhibiting Toll-like receptors 2 and 4, competing directly with pathogens, synthesizing antimicrobial substances or other metabolites, improving intestinal barrier function, and/or favorably altering the gut microbiota and the brain-gut axis [184,185,186]. It is also thought that lactobacilli such as *Lactobacillus paracasei, Lactobacillus plantarum* and *Lactobacillus rhamnosus* have the capacity to bind to and inactivate some viruses via adsorptive and/or trapping mechanisms [187]. Other probiotic lactobacilli and even non-viable bacterium-like particles have been shown to reduce both viral attachment to host cells and viral titers, along with reducing cytokine synthesis, enhancing the antiviral IFN-α response, and inducing various other antiviral mechanisms [187,188,189,190,191,192,193,194,195]. These antiviral and immunobiotic mechanisms and others have been reviewed in detail elsewhere [31,182,196]. However, there is also a bi-directional relationship between the lungs and gut microbiota known as the gut-lung axis [197], whereby gut microbial metabolites and endotoxins may affect the lungs via the circulatory system and the lung microbiota in return may affect the gut [198]. Therefore, the gut-lung axis may play role in our future understanding of COVID-19 pathogenesis and become a target for probiotic treatments [199]. Moreover, as microbial dysbiosis of the respiratory tract and gut may play a role in some viral infections, it has been suggested that SARS-CoV-2 may interact with our commensal microbiota [31; 200; 10.3389/fmicb.2020.01840] and that the lung microbiome could play a role in developing immunity to viral infections [201]. These postulations, if correct, could lead to the development of novel probiotic and prebiotic treatments. However, significant research is required to confirm these associations and their relevance to patient care, if any.

Probiotic therapies and prophylactics may also confer some advantages for managing symptoms of COVID-19 or risks associated with its treatment. Probiotics have tentatively been associated with the reduction of risk and duration of viral upper respiratory tract infections [202,203,204]. Some meta-analyses that have assessed the efficacy of probiotics in viral respiratory infections have reported moderate reductions in the incidence and duration of infection [203,205]. Indeed, randomized controlled trials have shown that administering *Bacillus subtilis* and *Enterococcus faecalis* [206], *Lactobacillus rhamnosus GG* [207], or *Lactobacillus casei* and *Bifidobacterium breve* with galactooligosaccharides [208] via the nasogastric tube to ventilated patients reduced the occurrence of ventilator-associated pneumonia in comparison to the respective control groups in studies of viral infections and sepsis. These findings were also supported by a recent meta-analysis [209]. Additionally, COVID-19 patients carry a significant risk of ventilator-associated bacterial pneumonia [210], but it can be challenging for clinicians to diagnose this infection due to the fact that severe COVID-19 infection presents with the symptoms of pneumonia [211]. Therefore, an effective prophylactic therapy for ventilator-associated pneumonia in severe COVID-19 patients would carry significant therapeutic value. Additionally, in recent years, probiotics have become almost synonymous with the treatment of gastrointestinal issues due to their supposed anti-inflammatory and immunomodulatory effects [212]. Notably, gastrointestinal symptoms commonly occur in COVID-19 patients [213], and angiotensin-converting enzyme 2, the portal by which SARS-CoV-2 enters human cells, is highly expressed in enterocytes of the ileum and colon, suggesting that these organs may be a potential route of infection [214,215]. Indeed, SARS-CoV-2 viral RNA has been detected in human feces [216,217], and fecal-oral transmission of the virus has not yet been ruled out [218]. Rectal swabs of some SARS-CoV-2 positive pediatric patients persistently tested positive for several days despite negative nasopharyngeal tests, indicating the potential for fecal viral shedding [219]. However, there is conflicting evidence for the therapeutic value of various probiotics against the incidence or severity of gastrointestinal symptoms in viral or bacterial infections such as gastroenteritis [220,221]. Nevertheless, it has been proposed that the administration of probiotics to COVID-19 patients and healthcare workers may prevent or ameliorate the gastrointestinal symptoms of COVID-19, a hypothesis that several clinical trials are now preparing to investigate [222,223]. Other studies are investigating whether probiotics may affect patient outcomes following SARS-CoV-2 infection [224].

Generally, the efficacy of probiotic use is a controversial topic among scientists. In Europe, EFSA has banned the term probiotics on products labels, which has elicited either criticism for EFSA or support for probiotics from researchers in the field [179,225,226]. This regulation is due to the hyperbolic claims placed on the labels of various probiotic products, which lack rigorous scientific data to support their efficacy. Overall, the data supporting probiotics in the treatment or prevention of many different disorders and diseases is not conclusive, as the quality of the evidence is generally considered low [202]. However, in the case of probiotics and respiratory infections, the evidence seems to be supportive of their potential therapeutic value. Consequently, several investigations are underway to investigate the prophylactic and therapeutic potential of probiotics for COVID-19. The blind use of conventional probiotics for COVID-19 is currently cautioned against until the pathogenesis of SARS-CoV-2 can be further established [227]. Until clinical trials investigating the prophylactic and therapeutic potential of probiotics for COVID-19 are complete, it is not possible to provide an evidence-based recommendation for their use. Despite these concerns, complementary use of probiotics as an adjuvant therapeutic has been proposed by the Chinese National Health Commission and National Administration of Traditional Chinese Medicine [228]. While supply issues prevented the probiotics market from showing the same rapid response to the COVID-19 as some other supplements, many suppliers are reporting growth during the pandemic [229]. Therefore, the public response once again seems to have adopted supplements promoted as bolstering the immune response despite a lack of evidence suggesting they are beneficial for preventing or mitigating COVID-19.

## Discussion

In this review, we report the findings to date of analyses of several dietary supplements and nutraceuticals. While existing evidence suggests potential benefits of n-3 PUFA and probiotic supplementation for COVID-19 treatment and prophylaxis, clinical data is still lacking, although trials are underway. Both zinc and vitamin C supplementation in hospitalized patients seem to be associated with positive outcomes; however, further clinical trials are required. In any case, both vitamin C and zinc intake are part of a healthy diet and both likely presents minimal risk when supplemented for, though their potential prophylactic or therapeutic effects against COVID-19 are yet to be determined. On the other hand, mounting evidence from observational studies indicates that there is an association between vitamin D deficiency and COVID-19 incidence has also been supported by meta-analysis [230]. Indeed, scientists are working to confirm these findings and to determine whether a patient's serum 25-hydroxyvitamin D levels are also associated with COVID-19 severity. Clinical trials are required to determine whether preemptive vitamin D supplementation may mitigate against severe COVID-19. In terms of the therapeutic potential of vitamin D, initial evidence from clinical trials are conflicting, but seem to indicate that vitamin D supplementation may reduce COVID-19 severity [158]. The various clinical trials currently underway will be imperative to provide information on the efficacious use of vitamin D supplementation for COVID-19 prevention and/or treatment.

The purported prophylactic and therapeutic benefits of dietary supplements and nutraceuticals for multiple disorders, diseases, and infections has been the subject of significant research and debate for the last few decades. Inevitably, scientists are also investigating the potential for these various products to treat or prevent COVID-19. This interest also extends to consumers, which led to a remarkable increase of sales of dietary supplements and nutraceuticals throughout the pandemic due to a desire to obtain additional protections from infection and disease. The nutraceuticals discussed in this review, namely vitamin C, vitamin D, n-3 PUFA, zinc, and probiotics, were selected because of potential biological mechanisms that could beneficially affect viral and respiratory infections and because they are currently under clinical investigation. Specifically, these compounds have all been found to influence cellular processes related to inflammation. Inflammation is particularly relevant to COVID-19 because of the negative outcomes (often death) observed in a large number of patients whose immune response becomes hyperactive in response to SARS-CoV-2, leading to severe outcomes such as ARDS and sepsis [231]. Additionally, there is a well-established link between diet and inflammation [232], potentially mediated in part by the microbiome [233]. Thus, the idea that dietary modifications or supplementation could be used to modify the inflammatory response is tied to a broader view of how diet and the immune system are interconnected. The supplements and nutraceuticals discussed here therefore lie in sharp contrast to other alleged nutraceutical or dietary supplements that have attracted during the pandemic, such as colloidal silver [234], which have no known nutritional function and can be harmful. Importantly, while little clinical evidence is available about the effects of any supplements against COVID-19, the risks associated with those discussed above are likely to be low, and in some cases, they can be obtained from dietary sources alone.

There are various other products and molecules that have garnered scientific interest and could merit further investigation. These include polyphenols, lipid extracts, and tomato-based nutraceuticals, all of which have been suggested for the potential prevention of cardiovascular complications of COVID-19 such as thrombosis [31,68]. Melatonin is another supplement that has been identified as a potential antiviral agent against SARS-CoV-2 using computational methods [235], and it has also been highlighted as a potential therapeutic agent for COVID-19 due to its documented antioxidant, anti-apoptotic, immunomodulatory, and anti-inflammatory effects [68,236,237]. Notably, melatonin, vitamin D and zinc have attracted public attention because they were included in the treatment plan of the former President of the United States upon his hospitalization due to COVID-19 [238]. These are just some of the many substances and supplements that are currently under investigation but as of yet lack evidence to support their use for the prevention or treatment of COVID-19. While there is plenty of skepticism put forward by physicians and scientists surrounding the use of supplements, these statements have not stopped consumers from purchasing these products, with one study reporting that online searches for dietary supplements in Poland began trending with the start of the pandemic [177]. Additionally, supplement usage increased between the first and second wave of the pandemic. Participants reported various reasons for their use of supplements, including to improve immunity (60%), to improve overall health (57%), and to fill nutrient gaps in their diet (53%). Other efforts to collect large datasets regarding such behavior have also sought to explore a possible association between vitamin or supplement consumption and COVID-19. An observational analysis of survey responses from 327,720 users of the COVID Symptom Study App found that the consumption of n-3 PUFA supplements, probiotics, multivitamins, and vitamin D was associated with a lower risk of SARS-CoV-2 infection in women but not men after adjusting for potential confounders [239]. According to the authors, the sexual dimorphism observed may in part be because supplements may better support females due to known differences between the male and female immune systems, or it could be due to behavioral and health consciousness differences between the sexes [239]. Certainly, randomized controlled trials are required to investigate these findings further.

Finally, it is known that a patient's nutritional status affects health outcomes in various infectious diseases [5], and COVID-19 is no different [3,240,241]. Some of the main risk factors for severe COVID-19, which also happen to be linked to poor nutritional status, include obesity, hypertension, cardiovascular diseases, type II diabetes mellitus, and indeed age-related malnutrition [1,3,242]. Although not the main focus of this review, it is important to consider the nutritional challenges associated with severe COVID-19 patients. Hospitalized COVID-19 patients tend to report an unusually high loss of appetite preceding admission, some suffer diarrhea and gastrointestinal symptoms that result in significantly lower food intake, and patients with poorer nutritional status were more likely to have worse outcomes and require nutrition therapy [243]. Dysphagia also seems to be a significant problem in pediatric patients that suffered multisystem inflammatory syndrome [244] and rehabilitating COVID-19 patients, potentially contributing to poor nutritional status [245]. Almost two-thirds of discharged COVID-19 ICU patients exhibit significant weight loss, of which 26% had weight loss greater than 10% [241]. As investigated in this review, hospitalized patients also tend to exhibit vitamin D deficiency or insufficiency, which may be associated with greater disease severity [230]. Therefore, further research is required to determine how dietary supplements and nutraceuticals may contribute to the treatment of severely ill and rehabilitating patients, who often rely on enteral nutrition.

## Conclusions

Despite all the potential benefits of nutraceutical and dietary supplement interventions presented, currently there is a paucity of clinical evidence to support their use for the prevention or mitigation of COVID-19 infection. Nevertheless, optimal nutritional status can prime an individual's immune system to protect against the effects of acute respiratory viral infections by supporting normal maintenance of the immune system [1,5]. Nutritional strategies can also play a role in the treatment of hospitalized patients, as malnutrition is a risk to COVID-19 patients [245]. Overall, supplementation of vitamin C, vitamin D, and zinc may be an effective method of ensuring their adequate intake to maintain optimal immune function, which may also convey beneficial effects against viral infections due to their immunomodulatory effects. Individuals should pay attention to their nutritional status, particularly their intake of vitamin D, considering that vitamin D deficiency is widespread. The prevailing evidence seems to indicate an association between vitamin D deficiency with COVID-19 incidence and, potentially, severity [133]. As a result, some international authorities have advised the general public, particularly those at high risk of infection, to consider vitamin D supplementation. However, further well-controlled clinical trials are required to confirm these observations.

Many supplements and nutraceuticals designed for various ailments that are available in the United States and beyond are not strictly regulated [246]. Consequently, there can be safety and efficacy concerns associated with many of these products. Often, the vulnerable members of society can be exploited in this regard and, unfortunately, the COVID-19 pandemic has proven no different. As mentioned above, the FDA has issued warnings to several companies for advertising falsified claims in relation to the preventative and therapeutic capabilities of their products against COVID-19 [247]. Further intensive investigation is required to establish the effects of these nutraceuticals, if any, against COVID-19. Until more effective therapeutics are established, the most effective mitigation strategies consist of encouraging standard public health practices such as regular hand washing with soap, wearing a face mask, and covering a cough with your elbow [248], along with following social distancing measures, "stay at home" guidelines, expansive testing, and contact tracing [249,250]. Indeed, in light of this review, it would also be pertinent to adopt a healthy diet and lifestyle following national guidelines in order to maintain optimal immune health. Because of the broad public appeal of dietary supplements and nutraceuticals, it is important to evaluate the evidence regarding the use of such products. We will continue to update this review as more findings become available.

# Additional Items

## Competing Interests

| Author | Competing Interests | Last Reviewed |
|---|---|---|
| Ronan Lordan | None | 2020-11-03 |
| Halie M. Rando | None | 2021-01-20 |
| COVID-19 Review Consortium | None | 2021-01-16 |
| Casey S. Greene | None | 2021-01-20 |

## Author Contributions

| Author | Contributions |
|---|---|
| Ronan Lordan | Conceptualization, Project Administration, Writing - Original Draft, Writing - Review & Editing |
| Halie M. Rando | Writing - Original Draft, Writing - Review & Editing |
| COVID-19 Review Consortium | Project Administration |
| Casey S. Greene | Project Administration, Supervision |

## Acknowledgements

We thank Nick DeVito for assistance with the Evidence-Based Medicine Data Lab COVID-19 TrialsTracker data and Vincent Rubinetti and Daniel Himmelstein for feedback on and support with Manubot. We thank Yael Evelyn Marshall who contributed writing (original draft) as well as reviewing and editing of pieces of the text but who did not formally approve the manuscript, as well as Ronnie Russell, who contributed text to and helped develop the structure of the manuscript early in the writing process and Matthias Fax who helped with writing and editing text related to diagnostics. We are grateful to the following contributors for reviewing pieces of the text: Nadia Danilova, James Eberwine and Ipsita Krishnan.

2. **Could nutrition modulate COVID-19 susceptibility and severity of disease? A systematic review**
Philip T. James, Zakari Ali, Andrew E. Armitage, Ana Bonell, Carla Cerami, Hal Drakesmith, Modou Jobe, Kerry S. Jones, Zara Liew, Sophie E. Moore, … Andrew M. Prentice
*Cold Spring Harbor Laboratory* (2020-10-21) https://doi.org/ghr94g
DOI: 10.1101/2020.10.19.20214395

3. **Coronavirus Disease 2019 (COVID-19) and Nutritional Status: The Missing Link?**
Renata Silverio, Daniela Caetano Gonçalves, Márcia Fábia Andrade, Marilia Seelaender
*Advances in Nutrition* (2020-09-25) https://doi.org/ghhqjd
DOI: 10.1093/advances/nmaa125 · PMID: 32975565 · PMCID: PMC7543263

4. **Nutritional status of patients with COVID-19**
Jae Hyoung Im, Young Soo Je, Jihyeon Baek, Moon-Hyun Chung, Hea Yoon Kwon, Jin-Soo Lee
*International Journal of Infectious Diseases* (2020-11) https://doi.org/gg7t5t
DOI: 10.1016/j.ijid.2020.08.018 · PMID: 32795605 · PMCID: PMC7418699

5. **Optimal Nutritional Status for a Well-Functioning Immune System Is an Important Factor to Protect against Viral Infections**
Philip C. Calder, Anitra C. Carr, Adrian F. Gombart, Manfred Eggersdorfer
*Nutrients* (2020-04-23) https://doi.org/gg29hh
DOI: 10.3390/nu12041181 · PMID: 32340216 · PMCID: PMC7230749

6. **Peak dietary supplement sales leveling off during COVID-19 pandemic, but growth still remains strong over last year, market researchers report during webcast**
Nutritional Outlook
https://www.nutritionaloutlook.com/view/peak-dietary-supplement-sales-leveling-during-covid-19-pandemic-growth-still-remains-strong

7. **Dietary Diversity among Chinese Residents during the COVID-19 Outbreak and Its Associated Factors**
Ai Zhao, Zhongyu Li, Yalei Ke, Shanshan Huo, Yidi Ma, Yumei Zhang, Jian Zhang, Zhongxia Ren
*Nutrients* (2020-06-06) https://doi.org/ghc6d9
DOI: 10.3390/nu12061699 · PMID: 32517210 · PMCID: PMC7352896

8. **Lockdown impact: Grocery stores bolstered NZ supplements sales as pharmacies slumped**
nutraingredients-asia.com
*nutraingredients-asia.com* https://www.nutraingredients-asia.com/Article/2020/07/06/Lockdown-impact-Grocery-stores-bolstered-NZ-supplements-sales-as-pharmacies-slumped

9. **COVID-19 temporarily bolsters European interest in supplements**
.nutritioninsight.com/
https://ni.cnsmedia.com/a/EHHJsDOG2oc=

70. **Platelet activation and prothrombotic mediators at the nexus of inflammation and atherosclerosis: Potential role of antiplatelet agents**
Ronan Lordan, Alexandros Tsoupras, Ioannis Zabetakis
*Blood Reviews* (2020-04) https://doi.org/ggvv7x
DOI: 10.1016/j.blre.2020.100694 · PMID: 32340775

71. **An Investigation on the Effects of Icosapent Ethyl (VascepaTM) on Inflammatory Biomarkers in Individuals With COVID-19 - Full Text View - ClinicalTrials.gov**
https://clinicaltrials.gov/ct2/show/NCT04412018

72. **Cardiovascular Risk Reduction with Icosapent Ethyl for Hypertriglyceridemia**
Deepak L. Bhatt, P. Gabriel Steg, Michael Miller, Eliot A. Brinton, Terry A. Jacobson, Steven B. Ketchum, Ralph T. Doyle, Rebecca A. Juliano, Lixia Jiao, Craig Granowitz, … Christie M. Ballantyne
*New England Journal of Medicine* (2019-01-03) https://doi.org/gfj3w9
DOI: 10.1056/nejmoa1812792 · PMID: 30415628

73. **A Randomised, Double-blind, Placebo Controlled Study of Eicosapentaenoic Acid (EPA-FFA) Gastro-resistant Capsules to Treat Hospitalised Subjects With Confirmed SARS-CoV-2**
S.L.A. Pharma AG
*clinicaltrials.gov* (2020-10-29) https://clinicaltrials.gov/ct2/show/NCT04335032

74. **Anti-inflammatory/Antioxidant Oral Nutrition Supplementation on the Cytokine Storm and Progression of COVID-19: A Randomized Controlled Trial**
Mahmoud Abulmeaty FACN M. D.
*clinicaltrials.gov* (2020-09-18) https://clinicaltrials.gov/ct2/show/NCT04323228

75. **Functional Role of Dietary Intervention to Improve the Outcome of COVID-19: A Hypothesis of Work**
Giovanni Messina, Rita Polito, Vincenzo Monda, Luigi Cipolloni, Nunzio Di Nunno, Giulio Di Mizio, Paolo Murabito, Marco Carotenuto, Antonietta Messina, Daniela Pisanelli, … Francesco Sessa
*International Journal of Molecular Sciences* (2020-04-28) https://doi.org/ggvb88
DOI: 10.3390/ijms21093104 · PMID: 32354030 · PMCID: PMC7247152

76. **Zinc and immunity: An essential interrelation**
Maria Maares, Hajo Haase
*Archives of Biochemistry and Biophysics* (2016-12) https://doi.org/f9c9b5
DOI: 10.1016/j.abb.2016.03.022 · PMID: 27021581

77. **Zinc-Dependent Suppression of TNF-α Production Is Mediated by Protein Kinase A-Induced Inhibition of Raf-1, IκB Kinase β, and NF-κB**
Verena von Bülow, Svenja Dubben, Gabriela Engelhardt, Silke Hebel, Birgit Plümäkers, Holger Heine, Lothar Rink, Hajo Haase
*The Journal of Immunology* (2007-09-15) https://doi.org/f3vs45
DOI: 10.4049/jimmunol.179.6.4180 · PMID: 17785857

86. **The SARS-coronavirus papain-like protease: Structure, function and inhibition by designed antiviral compounds**
Yahira M. Báez-Santos, Sarah E. St. John, Andrew D. Mesecar
*Antiviral Research* (2015-03) https://doi.org/f63hjp
DOI: 10.1016/j.antiviral.2014.12.015 · PMID: 25554382 · PMCID: PMC5896749

87. **A Randomized Study Evaluating the Safety and Efficacy of Hydroxychloroquine and Zinc in Combination With Either Azithromycin or Doxycycline for the Treatment of COVID-19 in the Outpatient Setting**
Avni Thakore MD
*clinicaltrials.gov* (2020-12-08) https://clinicaltrials.gov/ct2/show/NCT04370782

88. **A Study of Hydroxychloroquine and Zinc in the Prevention of COVID-19 Infection in Military Healthcare Workers - Full Text View - ClinicalTrials.gov**
https://clinicaltrials.gov/ct2/show/NCT04377646

89. **Therapies to Prevent Progression of COVID-19, Including Hydroxychloroquine, Azithromycin, Zinc, Vitamin D, Vitamin B12 With or Without Vitamin C, a Multi-centre, International, Randomized Trial: The International ALLIANCE Study**
National Institute of Integrative Medicine, Australia
*clinicaltrials.gov* (2020-09-09) https://clinicaltrials.gov/ct2/show/NCT04395768

90. **Early Intervention in COVID-19: Favipiravir Verses Standard Care - Full Text View - ClinicalTrials.gov** https://clinicaltrials.gov/ct2/show/NCT04373733

91. **Hydroxychloroquine with or without Azithromycin in Mild-to-Moderate Covid-19**
Alexandre B. Cavalcanti, Fernando G. Zampieri, Regis G. Rosa, Luciano C. P. Azevedo, Viviane C. Veiga, Alvaro Avezum, Lucas P. Damiani, Aline Marcadenti, Letícia Kawano-Dourado, Thiago Lisboa, … Otavio Berwanger
*New England Journal of Medicine* (2020-11-19) https://doi.org/gg5343
DOI: 10.1056/nejmoa2019014 · PMID: 32706953 · PMCID: PMC7397242

92. **The efficacy and safety of hydroxychloroquine for COVID-19 prophylaxis: A systematic review and meta-analysis of randomized trials**
Kimberley Lewis, Dipayan Chaudhuri, Fayez Alshamsi, Laiya Carayannopoulos, Karin Dearness, Zain Chagla, Waleed Alhazzani, for the GUIDE Group
*PLOS ONE* (2021-01-06) https://doi.org/ghsv36
DOI: 10.1371/journal.pone.0244778 · PMID: 33406138 · PMCID: PMC7787432

93. **Effect of hydroxychloroquine with or without azithromycin on the mortality of coronavirus disease 2019 (COVID-19) patients: a systematic review and meta-analysis**
Thibault Fiolet, Anthony Guihur, Mathieu Edouard Rebeaud, Matthieu Mulot, Nathan Peiffer-Smadja, Yahya Mahamat-Saleh
*Clinical Microbiology and Infection* (2021-01) https://doi.org/gg9jk2
DOI: 10.1016/j.cmi.2020.08.022 · PMID: 32860962 · PMCID: PMC7449662

94. **Zinc sulfate in combination with a zinc ionophore may improve outcomes in hospitalized COVID-19 patients**
Philip M. Carlucci, Tania Ahuja, Christopher Petrilli, Harish Rajagopalan, Simon Jones, Joseph Rahimian
*Journal of Medical Microbiology* (2020-10-01) https://doi.org/ghnws7
DOI: 10.1099/jmm.0.001250 · PMID: 32930657 · PMCID: PMC7660893

95. **The Minimal Effect of Zinc on the Survival of Hospitalized Patients With COVID-19**
Jasper Seth Yao, Joseph Alexander Paguio, Edward Christopher Dee, Hanna Clementine Tan, Achintya Moulick, Carmelo Milazzo, Jerry Jurado, Nicolás Della Penna, Leo Anthony Celi
*Chest* (2021-01) https://doi.org/gg5w36
DOI: 10.1016/j.chest.2020.06.082 · PMID: 32710890 · PMCID: PMC7375307

96. **Coronavirus Disease 2019- Using Ascorbic Acid and Zinc Supplementation (COVIDAtoZ) Research Study A Randomized, Open Label Single Center Study**
Milind Desai
*clinicaltrials.gov* (2021-01-28) https://clinicaltrials.gov/ct2/show/NCT04342728

97. **Vitamin B12 May Inhibit RNA-Dependent-RNA Polymerase Activity of nsp12 from the COVID-19 Virus**
Naveen Narayanan, Deepak T. Nair
*Preprints* (2020-03-22) https://doi.org/ggqmjc
DOI: 10.20944/preprints202003.0347.v1

98. **The Long History of Vitamin C: From Prevention of the Common Cold to Potential Aid in the Treatment of COVID-19**
Giuseppe Cerullo, Massimo Negro, Mauro Parimbelli, Michela Pecoraro, Simone Perna, Giorgio Liguori, Mariangela Rondanelli, Hellas Cena, Giuseppe D'Antona
*Frontiers in Immunology* (2020-10-28) https://doi.org/ghr943
DOI: 10.3389/fimmu.2020.574029 · PMID: 33193359 · PMCID: PMC7655735

99. **The Emerging Role of Vitamin C in the Prevention and Treatment of COVID-19**
Anitra C. Carr, Sam Rowe
*Nutrients* (2020-10-27) https://doi.org/ghr95c
DOI: 10.3390/nu12113286 · PMID: 33121019 · PMCID: PMC7693980

100. **Vitamin C Mitigates Oxidative Stress and Tumor Necrosis Factor-Alpha in Severe Community-Acquired Pneumonia and LPS-Induced Macrophages**
Yuanyuan Chen, Guangyan Luo, Jiao Yuan, Yuanyuan Wang, Xiaoqiong Yang, Xiaoyun Wang, Guoping Li, Zhiguang Liu, Nanshan Zhong
*Mediators of Inflammation* (2014) https://doi.org/f6nb5f
DOI: 10.1155/2014/426740 · PMID: 25253919 · PMCID: PMC4165740

101. **Intravenous infusion of ascorbic acid decreases serum histamine concentrations in patients with allergic and non-allergic diseases**
Alexander F. Hagel, Christian M. Layritz, Wolfgang H. Hagel, Hans-Jürgen Hagel, Edith Hagel, Wolfgang Dauth, Jürgen Kressel, Tanja Regnet, Andreas Rosenberg, Markus F. Neurath, …

134. **COVID-19 fatalities, latitude, sunlight, and vitamin D**
Paul B. Whittemore
*American Journal of Infection Control* (2020-09) https://doi.org/ghr93r
DOI: 10.1016/j.ajic.2020.06.193 · PMID: 32599103 · PMCID: PMC7319635

135. **Editorial: low population mortality from COVID-19 in countries south of latitude 35 degrees North supports vitamin D as a factor determining severity**
Jonathan M. Rhodes, Sreedhar Subramanian, Eamon Laird, Rose A. Kenny
*Alimentary Pharmacology & Therapeutics* (2020-06) https://doi.org/ggtw4b
DOI: 10.1111/apt.15777 · PMID: 32311755 · PMCID: PMC7264531

136. **25-Hydroxyvitamin D Concentrations Are Lower in Patients with Positive PCR for SARS-CoV-2**
Antonio D'Avolio, Valeria Avataneo, Alessandra Manca, Jessica Cusato, Amedeo De Nicolò, Renzo Lucchini, Franco Keller, Marco Cantù
*Nutrients* (2020-05-09) https://doi.org/ggvv76
DOI: 10.3390/nu12051359 · PMID: 32397511 · PMCID: PMC7285131

137. **Vitamin D deficiency as risk factor for severe COVID-19: a convergence of two pandemics**
D. De Smet, K. De Smet, P. Herroelen, S. Gryspeerdt, G. A. Martens
*Cold Spring Harbor Laboratory* (2020-05-05) https://doi.org/ggvv75
DOI: 10.1101/2020.05.01.20079376

138. **Vitamin D sufficiency, a serum 25-hydroxyvitamin D at least 30 ng/mL reduced risk for adverse clinical outcomes in patients with COVID-19 infection**
Zhila Maghbooli, Mohammad Ali Sahraian, Mehdi Ebrahimi, Marzieh Pazoki, Samira Kafan, Hedieh Moradi Tabriz, Azar Hadadi, Mahnaz Montazeri, Mehrad Nasiri, Arash Shirvani, Michael F. Holick
*PLOS ONE* (2020-09-25) https://doi.org/ghdzx8
DOI: 10.1371/journal.pone.0239799 · PMID: 32976513 · PMCID: PMC7518605

139. **Role of vitamin D in preventing of COVID-19 infection, progression and severity**
Nurshad Ali
*Journal of Infection and Public Health* (2020-10) https://doi.org/ghdzw9
DOI: 10.1016/j.jiph.2020.06.021 · PMID: 32605780 · PMCID: PMC7305922

140. **Low plasma 25(OH) vitamin D level is associated with increased risk of COVID-19 infection: an Israeli population-based study**
Eugene Merzon, Dmitry Tworowski, Alessandro Gorohovski, Shlomo Vinker, Avivit Golan Cohen, Ilan Green, Milana Frenkel-Morgenstern
*The FEBS Journal* (2020-08-28) https://doi.org/gg7b5c
DOI: 10.1111/febs.15495 · PMID: 32700398 · PMCID: PMC7404739

141. **Association of Vitamin D Status and Other Clinical Characteristics With COVID-19 Test Results**
David O. Meltzer, Thomas J. Best, Hui Zhang, Tamara Vokes, Vineet Arora, Julian Solway

*JAMA Network Open* (2020-09-03) https://doi.org/ghdzw6
DOI: 10.1001/jamanetworkopen.2020.19722 · PMID: 32880651 · PMCID: PMC7489852

142. **Vitamin D Status in Hospitalized Patients with SARS-CoV-2 Infection**
José L Hernández, Daniel Nan, Marta Fernandez-Ayala, Mayte García-Unzueta, Miguel A Hernández-Hernández, Marcos López-Hoyos, Pedro Muñoz-Cacho, José M Olmos, Manuel Gutiérrez-Cuadra, Juan J Ruiz-Cubillán, … Víctor M Martínez-Taboada
*The Journal of Clinical Endocrinology & Metabolism* (2020-10-27) https://doi.org/ghh737
DOI: 10.1210/clinem/dgaa733 · PMID: 33159440

143. **Analysis of vitamin D level among asymptomatic and critically ill COVID-19 patients and its correlation with inflammatory markers**
Anshul Jain, Rachna Chaurasia, Narendra Singh Sengar, Mayank Singh, Sachin Mahor, Sumit Narain
*Scientific Reports* (2020-11-19) https://doi.org/ghm3zn
DOI: 10.1038/s41598-020-77093-z · PMID: 33214648 · PMCID: PMC7677378

144. **Low 25-Hydroxyvitamin D Levels on Admission to the Intensive Care Unit May Predispose COVID-19 Pneumonia Patients to a Higher 28-Day Mortality Risk: A Pilot Study on a Greek ICU Cohort**
Alice G. Vassiliou, Edison Jahaj, Maria Pratikaki, Stylianos E. Orfanos, Ioanna Dimopoulou, Anastasia Kotanidou
*Nutrients* (2020-12-09) https://doi.org/ghr95d
DOI: 10.3390/nu12123773 · PMID: 33316914 · PMCID: PMC7764169

145. **Vitamin D deficiency as a predictor of poor prognosis in patients with acute respiratory failure due to COVID-19**
G. E. Carpagnano, V. Di Lecce, V. N. Quaranta, A. Zito, E. Buonamico, E. Capozza, A. Palumbo, G. Di Gioia, V. N. Valerio, O. Resta
*Journal of Endocrinological Investigation* (2020-08-09) https://doi.org/gg7kqp
DOI: 10.1007/s40618-020-01370-x · PMID: 32772324 · PMCID: PMC7415009

146. **Vitamin D Deficiency and Outcome of COVID-19 Patients**
Aleksandar Radujkovic, Theresa Hippchen, Shilpa Tiwari-Heckler, Saida Dreher, Monica Boxberger, Uta Merle
*Nutrients* (2020-09-10) https://doi.org/ghgfmp
DOI: 10.3390/nu12092757 · PMID: 32927735 · PMCID: PMC7551780

147. **Impact of Vitamin D Deficiency on COVID-19—A Prospective Analysis from the CovILD Registry**
Alex Pizzini, Magdalena Aichner, Sabina Sahanic, Anna Böhm, Alexander Egger, Gregor Hoermann, Katharina Kurz, Gerlig Widmann, Rosa Bellmann-Weiler, Günter Weiss, … Judith Löffler-Ragg
*Nutrients* (2020-09-11) https://doi.org/ghr95b
DOI: 10.3390/nu12092775 · PMID: 32932831 · PMCID: PMC7551662

148. **Does Serum Vitamin D Level Affect COVID-19 Infection and Its Severity?-A Case-Control Study**
Kun Ye, Fen Tang, Xin Liao, Benjamin A. Shaw, Meiqiu Deng, Guangyi Huang, Zhiqiang Qin, Xiaomei Peng, Hewei Xiao, Chunxia Chen, … Jianrong Yang
*Journal of the American College of Nutrition* (2020-10-13) https://doi.org/ghr935
DOI: 10.1080/07315724.2020.1826005 · PMID: 33048028

149. **Lower levels of vitamin D are associated with SARS-CoV-2 infection and mortality in the Indian population: An observational study**
Sunali Padhi, Subham Suvankar, Venketesh K. Panda, Abhijit Pati, Aditya K. Panda
*International Immunopharmacology* (2020-11) https://doi.org/ghr93w
DOI: 10.1016/j.intimp.2020.107001 · PMID: 33182040 · PMCID: PMC7489890

150. **Vitamin D Deficiency Is Associated with COVID-19 Incidence and Disease Severity in Chinese People**
Xia Luo, Qing Liao, Ying Shen, Huijun Li, Liming Cheng
*The Journal of Nutrition* (2021-01) https://doi.org/ghr939
DOI: 10.1093/jn/nxaa332 · PMID: 33188401

151. **Vitamin D concentrations and COVID-19 infection in UK Biobank**
Claire E. Hastie, Daniel F. Mackay, Frederick Ho, Carlos A. Celis-Morales, Srinivasa Vittal Katikireddi, Claire L. Niedzwiedz, Bhautesh D. Jani, Paul Welsh, Frances S. Mair, Stuart R. Gray, … Jill P. Pell
*Diabetes & Metabolic Syndrome: Clinical Research & Reviews* (2020-07) https://doi.org/ggvv72
DOI: 10.1016/j.dsx.2020.04.050 · PMID: 32413819 · PMCID: PMC7204679

152. **Vitamin D and COVID-19 infection and mortality in UK Biobank**
Claire E. Hastie, Jill P. Pell, Naveed Sattar
*European Journal of Nutrition* (2020-08-26) https://doi.org/ghr93p
DOI: 10.1007/s00394-020-02372-4 · PMID: 32851419 · PMCID: PMC7449523

153. **Low serum 25-hydroxyvitamin D (25[OH]D) levels in patients hospitalized with COVID-19 are associated with greater disease severity**
Grigorios Panagiotou, Su Ann Tee, Yasir Ihsan, Waseem Athar, Gabriella Marchitelli, Donna Kelly, Christopher S. Boot, Nadia Stock, James Macfarlane, Adrian R. Martineau, … Richard Quinton
*Clinical Endocrinology* (2020-08-06) https://doi.org/gg5gbj
DOI: 10.1111/cen.14276 · PMID: 32621392 · PMCID: PMC7361912

154. **Letter in response to the article: Vitamin D concentrations and COVID-19 infection in UK biobank (Hastie et al.)**
W. B. Grant, S. L. McDonnell
*Diabetes & Metabolic Syndrome: Clinical Research & Reviews* (2020-09) https://doi.org/ghc7p4
DOI: 10.1016/j.dsx.2020.05.046 · PMID: 32563941 · PMCID: PMC7293469

155. **Vitamin D deficiency in African Americans is associated with a high risk of severe disease and mortality by SARS-CoV-2**

Virna Margarita Martín Giménez, Felipe Inserra, León Ferder, Joxel García, Walter Manucha
*Journal of Human Hypertension* (2020-08-13) https://doi.org/ghr933
DOI: 10.1038/s41371-020-00398-z · PMID: 32792611 · PMCID: PMC7425793

156. **Evidence for possible association of vitamin D status with cytokine storm and unregulated inflammation in COVID-19 patients**
Ali Daneshkhah, Vasundhara Agrawal, Adam Eshein, Hariharan Subramanian, Hemant Kumar Roy, Vadim Backman
*Aging Clinical and Experimental Research* (2020-09-02) https://doi.org/ghr93q
DOI: 10.1007/s40520-020-01677-y · PMID: 32876941 · PMCID: PMC7465887

157. **Short term, high-dose vitamin D supplementation for COVID-19 disease: a randomised, placebo-controlled, study (SHADE study)**
Ashu Rastogi, Anil Bhansali, Niranjan Khare, Vikas Suri, Narayana Yaddanapudi, Naresh Sachdeva, GD Puri, Pankaj Malhotra
*Postgraduate Medical Journal* (2020-11-12) https://doi.org/ghnhpq
DOI: 10.1136/postgradmedj-2020-139065 · PMID: 33184146

158. **"Effect of calcifediol treatment and best available therapy versus best available therapy on intensive care unit admission and mortality among patients hospitalized for COVID-19: A pilot randomized clinical study"**
Marta Entrenas Castillo, Luis Manuel Entrenas Costa, José Manuel Vaquero Barrios, Juan Francisco Alcalá Díaz, José López Miranda, Roger Bouillon, José Manuel Quesada Gomez
*The Journal of Steroid Biochemistry and Molecular Biology* (2020-10) https://doi.org/ghd79r
DOI: 10.1016/j.jsbmb.2020.105751 · PMID: 32871238 · PMCID: PMC7456194

159. **COVID-19 rapid evidence summary: vitamin D for COVID-19 | Advice | NICE**
https://www.nice.org.uk/advice/es28

160. **Mathematical analysis of Córdoba calcifediol trial suggests strong role for Vitamin D in reducing ICU admissions of hospitalized COVID-19 patients**
Irwin Jungreis, Manolis Kellis
*Cold Spring Harbor Laboratory* (2020-12-21) https://doi.org/ghr94h
DOI: 10.1101/2020.11.08.20222638

161. **High-Dose Cholecalciferol Booster Therapy is Associated with a Reduced Risk of Mortality in Patients with COVID-19: A Cross-Sectional Multi-Centre Observational Study**
Stephanie F. Ling, Eleanor Broad, Rebecca Murphy, Joseph M. Pappachan, Satveer Pardesi-Newton, Marie-France Kong, Edward B. Jude
*Nutrients* (2020-12-11) https://doi.org/ghr95f
DOI: 10.3390/nu12123799 · PMID: 33322317 · PMCID: PMC7763301

162. **Effect of Vitamin D₃ Supplementation vs Placebo on Hospital Length of Stay in Patients with Severe COVID-19: A Multicenter, Double-blind, Randomized Controlled Trial**
Igor H. Murai, Alan L. Fernandes, Lucas P. Sales, Ana J. Pinto, Karla F. Goessler, Camila S. C. Duran, Carla B. R. Silva, André S. Franco, Marina B. Macedo, Henrique H. Dalmolin, … Rosa M. R. Pereira

*Cold Spring Harbor Laboratory* (2020-11-17) https://doi.org/ghr94j
DOI: 10.1101/2020.11.16.20232397

163. **Effect of Vitamin D Administration on Prevention and Treatment of Mild Forms of Suspected Covid-19**
Manuel Castillo Garzón
*clinicaltrials.gov* (2020-04-03) https://clinicaltrials.gov/ct2/show/NCT04334005

164. **Improving Vitamin D Status in the Management of COVID-19**
Aldo Montano-Loza
*clinicaltrials.gov* (2020-06-03) https://clinicaltrials.gov/ct2/show/NCT04385940

165. **Cholecalciferol to Improve the Outcomes of COVID-19 Patients - Full Text View - ClinicalTrials.gov** https://clinicaltrials.gov/ct2/show/NCT04411446

166. **COvid-19 and Vitamin D Supplementation: a Multicenter Randomized Controlled Trial of High Dose Versus Standard Dose Vitamin D3 in High-risk COVID-19 Patients (CoVitTrial) - Full Text View - ClinicalTrials.gov**
https://clinicaltrials.gov/ct2/show/NCT04344041

167. **The LEAD COVID-19 Trial: Low-risk, Early Aspirin and Vitamin D to Reduce COVID-19 Hospitalizations**
Louisiana State University Health Sciences Center in New Orleans
*clinicaltrials.gov* (2020-04-24) https://clinicaltrials.gov/ct2/show/NCT04363840

168. **Randomized Double-Blind Placebo-Controlled Proof-of-Concept Trial of a Plant Polyphenol for the Outpatient Treatment of Mild Coronavirus Disease (COVID-19)**
Marvin McCreary MD
*clinicaltrials.gov* (2020-09-22) https://clinicaltrials.gov/ct2/show/NCT04400890

169. **Current vitamin D status in European and Middle East countries and strategies to prevent vitamin D deficiency: a position statement of the European Calcified Tissue Society**
Paul Lips, Kevin D Cashman, Christel Lamberg-Allardt, Heike Annette Bischoff-Ferrari, Barbara Obermayer-Pietsch, Maria Luisa Bianchi, Jan Stepan, Ghada El-Hajj Fuleihan, Roger Bouillon
*European Journal of Endocrinology* (2019-04) https://doi.org/ggr42p
DOI: 10.1530/eje-18-0736 · PMID: 30721133

170. **Communiqué de l'Académie nationale de Médecine : Vitamine D et Covid-19 – Académie nationale de médecine | Une institution dans son temps** https://www.academie-medecine.fr/communique-de-lacademie-nationale-de-medecine-vitamine-d-et-covid-19/

171. **Covid-19: NHS bosses told to assess risk to ethnic minority staff who may be at greater risk**
Gareth Iacobucci
*BMJ* (2020-05-04) https://doi.org/ggv2zq
DOI: 10.1136/bmj.m1820 · PMID: 32366503

229. **Bloomberg - Are you a robot?**
https://www.bloomberg.com/tosv2.html?vid=&uuid=a9a62de0-6563-11eb-bdda-5f604b5cc36f&url=L3ByZXNzLXJlbGVhc2VzLlwMjAtMDgtMDMvcHJvYmlvdGljcy1tYXJrZXQtd29ydGgtNzYtNy1iaWxsaW9uLWJ5LTIwMjctZXhjbHVzaXZlLXJlcG9ydC1jb3ZlcmluZy1wcmUtYW5kLXBvc3QtY292aWQtMTktbWFya2V0LWFuYWx5c2lzLWJ5LW1tdGGljdWxvdXM=

230. **Vitamin D deficiency aggravates COVID-19: systematic review and meta-analysis**
Marcos Pereira, Alialdo Dantas Damascena, Laylla Mirella Galvão Azevedo, Tarcio de Almeida Oliveira, Jerusa da Mota Santana
*Critical Reviews in Food Science and Nutrition* (2020-11-04) https://doi.org/ghr937
DOI: 10.1080/10408398.2020.1841090 · PMID: 33146028

231. **Cytokine Storm**
David C. Fajgenbaum, Carl H. June
*New England Journal of Medicine* (2020-12-03) https://doi.org/ghnhm7
DOI: 10.1056/nejmra2026131 · PMID: 33264547 · PMCID: PMC7727315

232. **Diet and Inflammation**
Leo Galland
*Nutrition in Clinical Practice* (2010-12-07) https://doi.org/b7qgx7
DOI: 10.1177/0884533610385703 · PMID: 21139128

233. **Obesogenic diet in aging mice disrupts gut microbe composition and alters neutrophi:lymphocyte ratio, leading to inflamed milieu in acute heart failure**
Vasundhara Kain, William Van Der Pol, Nithya Mariappan, Aftab Ahmad, Peter Eipers, Deanna L. Gibson, Cecile Gladine, Claire Vigor, Thierry Durand, Casey Morrow, Ganesh V. Halade
*The FASEB Journal* (2019-02-15) https://doi.org/ghwfq8
DOI: 10.1096/fj.201802477r · PMID: 30768364 · PMCID: PMC6463911

234. **Colloidal Silver**
NCCIH
https://www.nccih.nih.gov/health/colloidal-silver

235. **Network-based drug repurposing for novel coronavirus 2019-nCoV/SARS-CoV-2**
Yadi Zhou, Yuan Hou, Jiayu Shen, Yin Huang, William Martin, Feixiong Cheng
*Cell Discovery* (2020-03-16) https://doi.org/ggq84x
DOI: 10.1038/s41421-020-0153-3 · PMID: 32194980 · PMCID: PMC7073332

236. **Role of Melatonin on Virus-Induced Neuropathogenesis—A Concomitant Therapeutic Strategy to Understand SARS-CoV-2 Infection**
Prapimpun Wongchitrat, Mayuri Shukla, Ramaswamy Sharma, Piyarat Govitrapong, Russel J. Reiter
*Antioxidants* (2021-01-02) https://doi.org/ghr946
DOI: 10.3390/antiox10010047 · PMID: 33401749 · PMCID: PMC7823793

237. **Nutraceutical Strategies for Suppressing NLRP3 Inflammasome Activation: Pertinence to the Management of COVID-19 and Beyond**

Mark F. McCarty, Simon Bernard Iloki Assanga, Lidianys Lewis Luján, James H. O'Keefe, James J. DiNicolantonio
*Nutrients* (2020-12-25) https://doi.org/ghr95g
DOI: 10.3390/nu13010047 · PMID: 33375692 · PMCID: PMC7823562

238. **Update: Here's what is known about Trump's COVID-19 treatment**
Jon Cohen
*Science* (2020-10-05) https://doi.org/ghr94n
DOI: 10.1126/science.abf0974

239. **Dietary supplements during the COVID-19 pandemic: insights from 1.4M users of the COVID Symptom Study app - a longitudinal app-based community survey**
Panayiotis Louca, Benjamin Murray, Kerstin Klaser, Mark S Graham, Mohsen Mazidi, Emily R Leeming, Ellen Thompson, Ruth Bowyer, David A Drew, Long H Nguyen, … Cristina Menni
*Cold Spring Harbor Laboratory* (2020-11-30) https://doi.org/ghr94k
DOI: 10.1101/2020.11.27.20239087

240. **ESPEN expert statements and practical guidance for nutritional management of individuals with SARS-CoV-2 infection**
Rocco Barazzoni, Stephan C. Bischoff, Joao Breda, Kremlin Wickramasinghe, Zeljko Krznaric, Dorit Nitzan, Matthias Pirlich, Pierre Singer
*Clinical Nutrition* (2020-06) https://doi.org/ggtzjq
DOI: 10.1016/j.clnu.2020.03.022 · PMID: 32305181 · PMCID: PMC7138149

241. **Nutritional status assessment in patients with Covid-19 after discharge from the intensive care unit**
Nassim Essabah Haraj, Siham El Aziz, Asma Chadli, Asma Dafir, Amal Mjabber, Ouissal Aissaoui, Lhoucine Barrou, Chafik El Kettani El Hamidi, Afak Nsiri, Rachid AL Harrar, … Moulay Hicham Afif
*Clinical Nutrition ESPEN* (2021-02) https://doi.org/ghjhdq
DOI: 10.1016/j.clnesp.2020.09.214 · PMID: 33487301 · PMCID: PMC7552965

242. **Nutrition Status Affects COVID‐19 Patient Outcomes**
Mette M Berger
*Journal of Parenteral and Enteral Nutrition* (2020-07-15) https://doi.org/gg5qv4
DOI: 10.1002/jpen.1954 · PMID: 32613691 · PMCID: PMC7361441

243. **Evaluation of Nutrition Risk and Its Association With Mortality Risk in Severely and Critically Ill COVID‐19 Patients**
Xiaobo Zhao, Yan Li, Yanyan Ge, Yuxin Shi, Ping Lv, Jianchu Zhang, Gui Fu, Yanfen Zhou, Ke Jiang, Nengxing Lin, … Xin Li
*Journal of Parenteral and Enteral Nutrition* (2020-07-20) https://doi.org/ghr93n
DOI: 10.1002/jpen.1953 · PMID: 32613660 · PMCID: PMC7361906

244. **Multisystem inflammatory syndrome in children: A systematic review**
Mubbasheer Ahmed, Shailesh Advani, Axel Moreira, Sarah Zoretic, John Martinez, Kevin Chorath, Sebastian Acosta, Rija Naqvi, Finn Burmeister-Morton, Fiona Burmeister, … Alvaro

Moreira
*EClinicalMedicine* (2020-09) https://doi.org/ghsv27
DOI: 10.1016/j.eclinm.2020.100527 · PMID: 32923992 · PMCID: PMC7473262

245. **Nutritional management of COVID-19 patients in a rehabilitation unit**
Luigia Brugliera, Alfio Spina, Paola Castellazzi, Paolo Cimino, Pietro Arcuri, Alessandra Negro, Elise Houdayer, Federica Alemanno, Alessandra Giordani, Pietro Mortini, Sandro Iannaccone
*European Journal of Clinical Nutrition* (2020-05-20) https://doi.org/gg29hf
DOI: 10.1038/s41430-020-0664-x · PMID: 32433599 · PMCID: PMC7237874

246. **The frontier between nutrition and pharma: The international regulatory framework of functional foods, food supplements and nutraceuticals**
Laura Domínguez Díaz, Virginia Fernández-Ruiz, Montaña Cámara
*Critical Reviews in Food Science and Nutrition* (2019-03-29) https://doi.org/ggqs3w
DOI: 10.1080/10408398.2019.1592107 · PMID: 30924346

247. **Coronavirus Update: FDA and FTC Warn Seven Companies Selling Fraudulent Products that Claim to Treat or Prevent COVID-19**
Office of the Commissioner
*FDA* (2020-03-27) https://www.fda.gov/news-events/press-announcements/coronavirus-update-fda-and-ftc-warn-seven-companies-selling-fraudulent-products-claim-treat-or

248. **COVID-19 and Your Health**
CDC
*Centers for Disease Control and Prevention* (2021-01-30)
https://www.cdc.gov/coronavirus/2019-ncov/prevent-getting-sick/prevention.html

249. **Potential roles of social distancing in mitigating the spread of coronavirus disease 2019 (COVID-19) in South Korea**
Sang Woo Park, Kaiyuan Sun, Cécile Viboud, Bryan T. Grenfell, Jonathan Dushoff
*Cold Spring Harbor Laboratory* (2020-03-30) https://doi.org/gg3mhg
DOI: 10.1101/2020.03.27.20045815 · PMID: 32511429 · PMCID: PMC7217070

250. **Evaluating the Effectiveness of Social Distancing Interventions to Delay or Flatten the Epidemic Curve of Coronavirus Disease**
Laura Matrajt, Tiffany Leung
*Emerging Infectious Diseases* (2020-08) https://doi.org/ggtx3k
DOI: 10.3201/eid2608.201093 · PMID: 32343222 · PMCID: PMC7392458